\documentclass[conference,a4paper,10pt]{IEEEtran}
\usepackage[latin1]{inputenc}
\usepackage{times,amsmath}
\usepackage{amssymb}
\usepackage{pstool}
\usepackage{subfigure}
\usepackage{caption}
\usepackage{multirow}
\usepackage{enumerate}
\usepackage{graphicx}
\usepackage{MnSymbol}
\usepackage{stfloats} 
\usepackage[table]{xcolor}
\usepackage[square, comma, sort&compress, numbers]{natbib}
\usepackage{nohyperref}
\usepackage{algorithm,algorithmic}
\usepackage{epstopdf}

\graphicspath{ {Figures/} }

\begin{document}

\title{Hybrid PLS-ML Authentication Scheme for V2I Communication Networks}

\author{
\IEEEauthorblockN{
 Hala Amin, Jawaher Kaldari, Nora Mohamed, Waqas Aman, Saif Al-Kuwari}
\IEEEauthorblockA{Division of Information and Computing Technology, College of Science and Engineering, \\Hamad Bin Khalifa University, Qatar Foundation, Doha, Qatar. \\
\{haam51711, jaka51804, nomo51812, waman, smalkuwari\}@hbku.edu.qa}
}

\maketitle
\long\def\symbolfootnote[#1]#2{\begingroup%
\def\thefootnote{\fnsymbol{footnote}}\footnote[#1]{#2}\endgroup}
\maketitle
\begin{abstract} 
Vehicular communication networks are rapidly emerging as vehicles become smarter. However, these networks are increasingly susceptible to various attacks. The situation is exacerbated by the rise in automated vehicles complicates, emphasizing the need for security and authentication measures to ensure safe and effective traffic management. In this paper, we propose a novel hybrid physical layer security (PLS)-machine learning (ML) authentication scheme by exploiting the position of the transmitter vehicle as a device fingerprint. We use a time-of-arrival (ToA) based localization mechanism where the ToA is estimated at roadside units (RSUs), and the coordinates of the transmitter vehicle are extracted at the base station (BS). 
Furthermore, to track the mobility of the moving legitimate vehicle, we use ML model trained on several system parameters.
We try two ML models for this purpose, i.e., support vector regression and decision tree. 

To evaluate our scheme, we conduct binary hypothesis testing on the estimated positions with the help of the ground truths provided by the ML model, which classifies the transmitter node as legitimate or malicious. Moreover, we consider the probability of false alarm and the probability of missed detection as performance metrics resulting from the binary hypothesis testing, and mean absolute error (MAE), mean square error (MSE), and coefficient of determination $\text{R}^2$ to further evaluate the ML models. We also compare our scheme with a baseline scheme that exploits angle of arrival at RSUs for authentication. We observe that our proposed position-based mechanism outperforms the baseline scheme significantly in terms of missed detections.   
 
\end{abstract}
\vspace{0.2cm}

\section{Introduction}
\label{sec:intro}
Vehicular communication networks (VCNs) are a type of communication system that enables wireless communication among vehicles and vehicles to roadside infrastructure. VCNs are designed to provide efficient and reliable communication in order to improve road safety, traffic efficiency, and the overall driving experience \cite{zeadally2020vehicular}.
There are two main types of VCNs: vehicle-to-vehicle (V2V) communication and vehicle-to-infrastructure (V2I) communication. V2V communication enables vehicles to communicate with each other and exchange information such as location, speed, and direction. This type of communication can be used to alert drivers to critical events on the road, such as an upcoming intersection or a vehicle stopped ahead \cite{bazzi2011taking}. V2I communication, on the other hand, enables vehicles to communicate with roadside infrastructure, such as traffic lights and sensors, in order to improve traffic flow and reduce congestion \cite{tahir2022connected}.







As a relatively recent type of networks, VCNs are vulnerable to cyberattacks, which can compromise the safety and privacy of drivers and passengers \cite{el2020cybersecurity}. Therefore, security in VCNs is crucial and needs to be ensured at the highest level. 
Authentication is one of the four main properties of security that need to be preserved in any secure system. Authentication verifies the identities of entities in VCNs, such as vehicles, infrastructure, and users before granting them access to the network. This helps to prevent unauthorized access and misuse of network resources. Authentication further provides secure access control, protects against impersonation attacks, safeguards sensitive data, and ensures trust in the system.
Generally, authentication schemes were mainly studied at the higher layer of protocol stacks where predefined secrets (i.e., passwords, keys, signatures) are utilized for this purpose. The secrets are encrypted and decrypted via various cryptographic measures \cite{alia2014cryptography}. However, some instances in the literature reported breaching cryptographic measures through brute force attacks \cite{gidney2021factor}. Therefore, alternative security mechanisms are now being evaluated. One such (promising) mechanism lies in the physical layer. Authentication at the physical layer is known as physical layer authentication (PLA) where the randomness in the characteristics of the physical layer is exploited. This randomness is mainly incurred in the wireless channel or hardware.  There are a variety of fingerprints/features exploited for PLA, including channel impulse response, channel frequency response, received signal strength indicator, transmission coefficient (S21), pathloss , I/Q imbalance, carrier offsets \cite{PLA:Network:2020,aman2023security}.
\vspace{-0.2cm}
\subsection{Related Work}
The authors in \cite{abdelaziz2018beyond} use the angle of arrival of the transmitter vehicle as a feature at the physical layer for authentication in the V2X environment. This work assumes that the location information of the transmitter node is available at the receiver and therefore expects a ground truth. This assumption is not realistic as there is no mechanism for ground truth tracking, and the effect of mobility is not discussed.  

Similarly, the authors in this paper \cite{9187413} proposed using physical layer characteristics for authentication and then using the Kalman filter to refine the iterative and threshold model. The iterative model estimates the priori and posteriori of the current time based on the physical layer characteristics of the previous time, serving as the basis for the authentication process. The threshold model analyzes the mathematical characteristics of the priori estimation and provides a calculation method for the authentication threshold. The authors also used an extended Kalman filter and unscented Kalman filter for nonlinear physical layer characteristics. 

In \cite{jadoon2021physical}, the authors proposed a novel authentication approach, referred to as Hopper-Blum based physical layer (HB-PL) authentication scheme, which incorporates an advanced physical layer key generation technique with the Hopper-Blum (HB) authentication scheme. In this scheme, information gathered from the shared channel is utilized as secret keys for the HB scheme, while the mismatched bits are applied as induced noise for solving the learning parity with noise (LPN) problem. The primary objective of the proposed technique is to offer a solution for the bit reconciliation process while ensuring that no information is exposed on a public channel. 

The work \cite{xu2022post} provides a PLA scheme that utilizes Gaussian process (GP) path loss prediction and channel state information (CSI) to track changes in channel characteristics and predict the next path loss (PL) of the signal from a transmitter for authentication. The scheme maps historical CSI attributes to PL features of the transmitter's signal to predict the next PL, which is then used to cross-verify the transmitter's reported location information \cite{umar2023physical}. 

More recently, the authors in \cite{shawky2023efficient} proposed a cross-layer authentication scheme for vehicular communication that uses short-term reciprocal features of the wireless channel to re-authenticate the corresponding terminal. By utilizing the reciprocal features of the wireless channel, the scheme aims to reduce the overall complexity and computation and communication overheads required for authentication.
\vspace{-0.2cm}
\subsection{Contribution}
In this work, we systematically adopt a novel approach exploiting the position of the transmitter node as a feature/fingerprint for authentication in V2I communication. Although position/location is very recently reported for PLA in underwater acoustic communication networks \cite{aman2023locationbased}, the scheme is limited to stationary nodes scenario. In this work, we assume a dynamic vehicular environment where vehicles are not stationary but moving at a certain speed. The main contributions of this work can be summarized as follows:
\begin{itemize}
    \item We estimate the position of the transmitter nodes by using  Time-of-Arrival (ToA) based localization method, where ToAs are estimated at the corresponding RSUs using the maximum likelihood approach and the coordinates are extracted at BS using the least square approach.
    \item We construct a test statistic on the extracted coordinates/estimated position for a binary hypothesis test to decide the legitimacy of the transmitter vehicle. To deal with the challenge of the mobility of the vehicle in hypothesis testing, we propose a machine-learning model to track the mobility of the legitimate node and predict the next position of the vehicle.
\end{itemize}
\vspace{-0.2cm}
\subsection{Organization}
The rest of this paper is structured as follows: In Section \ref{sec:SM}, we describe our system model. Section \ref{sec:PAS} presents the proposed physical layer authentication (PLA) scheme including position estimation, hypothesis testing, and the machine learning model. Section \ref{sec:simulation} presents the evaluation results of our proposed technique. Lastly, Section \ref{sec:conclusion} concludes the paper with a few final remarks and suggestions for future research directions.
\vspace{-0.2cm}

\section{System Model}
\label{sec:SM}
\begin{figure}[ht!]
    \centering
    \includegraphics[scale=0.3]{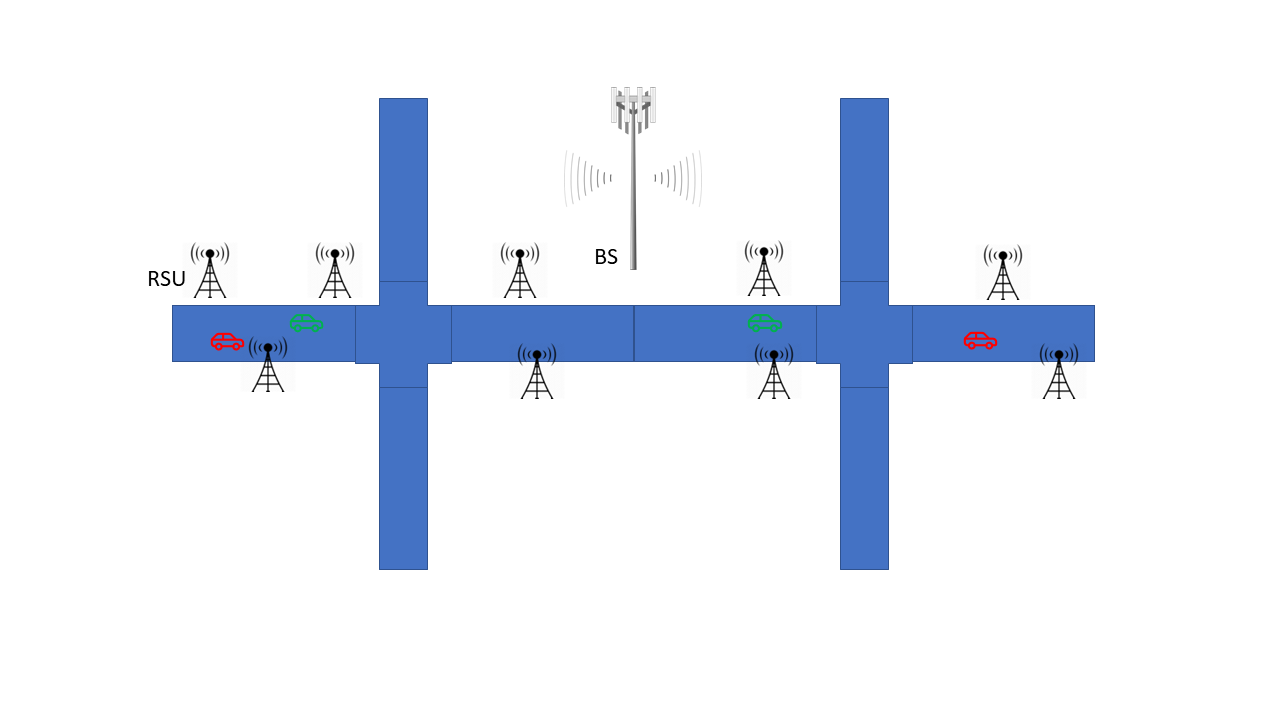}
    \caption{An Illustration of our System Model}
    \label{fig:SM}
\end{figure}

We assume the uplink transmission in a 2D V2I environment where vehicles are communicating with the RSUs to inform the central system (i.e., BS) about its parameters (speed, engine transmissions, fuel level, etc.) for congestion control or traffic management. We further assume that RSUs are deployed at fixed locations on both sides of the road and are connected to the BS as illustrated in Fig. \ref{fig:SM}.
We assume two kinds of vehicles: legitimate vehicles and malicious vehicles.  We assume a time-slotted communication system with no collision domain, i.e., only one transmitter node transmits at a given time slot. We assume that the malicious transmitter vehicle is smart enough and transmits in idle slots with the same transmit power as the legitimate vehicle so that she remains hidden in the network.  All the RSUs are assumed to be connected with BS via an error-free secured communication link. We assume the malicious vehicle attacks on the vehicles to RSUs links. Such attacks are often known as impersonation attacks.

\section{Proposed Authentication Scheme}
\label{sec:PAS}
\begin{figure}[htb!]
    \centering
    \includegraphics[scale=0.3]{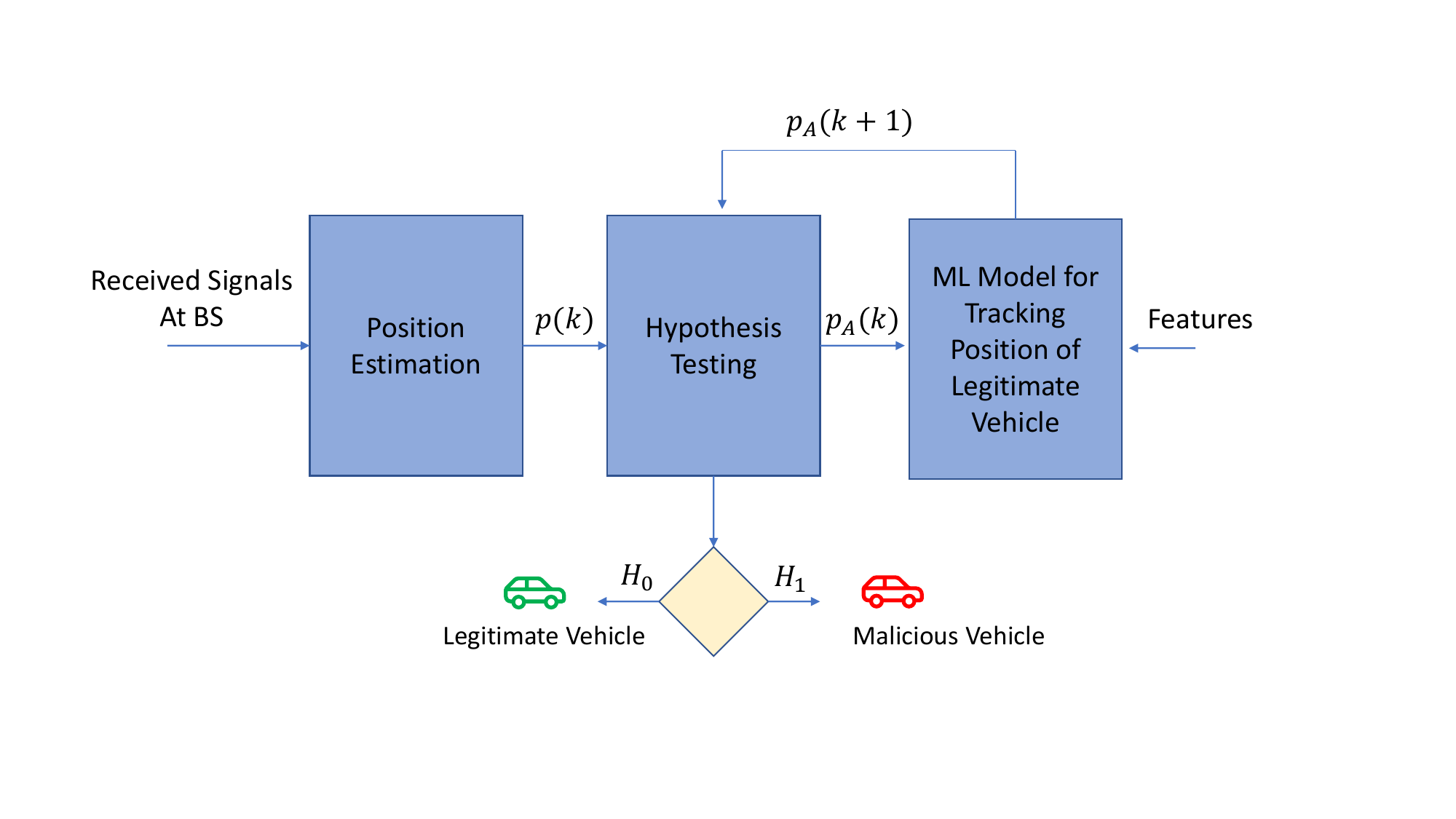}
    \caption{Proposed Methodology}
    \label{fig:PM}
\end{figure}

The proposed physical layer authentication scheme consists of three main components as depicted in Fig. \ref{fig:PM}. We discuss the functionality of each component in detail in the following subsections.

\subsection{Position Estimation}
The estimation of the position of the transmitter vehicle is accomplished in two stages: distance estimation and coordinates extraction.
\subsubsection{Distance Estimation}
The distances at RSUs are estimated from the ToAs. Let $\hat{t}_j$ be the estimated ToA at j-th RSU, which can be expressed as:\vspace{-0.2cm}
\begin{align}
    \hat{t}_{j}&= \underset{t_j}{\text{argmax}} \log f_\mathbf{y}(\mathbf{y}|t_j)  = \underset{t_j}{\text{argmax}} L(\mathbf{y}\mid t_j),
\end{align}

\noindent where $L(\mathbf{y}\mid t_j)$ is the log-likelihood function of the conditional random event $\mathbf{y}\mid t_j$ with $\mathbf{y}$ being the received symbols vector. According to the framework in \cite{Waqas:Access:2018}, $\hat{t}_j\sim \mathcal{N}(t_i,\sigma_t^2)$, where $\sigma_t^2=\frac{\sigma^2\psi_j}{4P}$ is the CRB or variance of the estimator with noise power $\sigma^2$, pathloss $\psi_j$ and transmit power $P$. Next, we use the famous distance equation, i.e., $\hat{r}_j=c\hat{t}_j$, to estimate the distance between the transmitter vehicle and j-th RSU, where $c=3\times 10^8$m/s is the speed of the RF-carrier. The estimated distance is distributed as $\hat{r}_j\sim \mathcal{N}(r_j,\sigma_r^2)$, where $r_j=ct_j$ is the actual distance and $\sigma_r^2=\frac{c^2\sigma^2\psi_j}{4P}$ is the variance of the distance estimator.

\subsubsection{Coordinates Extraction} 
Assuming $\mathbf{p}_j=\begin{bmatrix} x_j& y_j \end{bmatrix}^T $ is the position vector of the j-th RSU, and that $\mathbf{p}=\begin{bmatrix} x& y \end{bmatrix}^T $ is the unknown position vector or coordinates of the transmitter vehicle, the distance $r_j$ between the two nodes as per the definition of Euclidean distance is $r_j=\sqrt{(x-x_j)^2+(y-y_j)^2}$.
As ToA is susceptible to measurement error, the estimated measurement based on multiplying $v$ and $t
_j$ is denoted as $\hat{r}_j=r_j+n_j$. By squaring both sides, we get $\hat{r}_j^2=(r_j+n_j)^2=r_j^2+2n_jr_j+n_j^2$, which can be expressed as:
\begin{align}
\label{eq:seq}
\hat{r}_j^2=(x-x_j)^2+(y-y_j)^2+2n_j\sqrt{(x-x_j)^2+(y-y_j)^2}+n_j^2.
\end{align}
The equation set obtained from the Eq. \ref{eq:seq} can be expressed in a matrix-vector format for every instance of ''$j_s$'' as: \vspace{-0.3cm}
\begin{align}
\label{eq:matrix}
\mathbf{A\theta}+\mathbf{n}=\hat{\mathbf{b}}, 
\end{align}
where  all the vectors and matrices are given below: 
\begin{align}
&\mathbf{A}=
\begin{bmatrix}
    -2x_1 & -2y_1 & 1   \\
    . & . & .   \\
     . & . & .  \\
    -2x_L & -2y_L & 1      
  \end{bmatrix} , \ \hat{\mathbf{b}}=\begin{bmatrix}
      \hat{r}_1^2-x_1^2-y_1^2 \\
    .   \\
     .\\
       \hat{r}_L^2-x_L^2-y_L^2   
  \end{bmatrix} \, \mathbf{\theta}=\begin{bmatrix} 
    x   \\
    y  \\
     x^2+y^2     
  \end{bmatrix},   \nonumber \\ &\mathbf{b}=\begin{bmatrix}
      r_1^2-x_1^2-y_1^2 \\
    .   \\
     .\\
       r_L^2-x_L^2-y_L^2   
  \end{bmatrix},
\  \text{and} \ \mathbf{n}=
  \begin{bmatrix}
    2n_1\sqrt{(x-x_1)^2+(y-y_1)^2}+n_1^2   \\
    .    \\
    . \\
     2n_L\sqrt{(x-x_L)^2+(y-y_L)^2}+n_L^2     
  \end{bmatrix}.
 \nonumber 
\end{align}
Eq. \ref{eq:matrix} is a linear least square problem, where $\hat{\mathbf{b}}$ is the noisy observation vector.  The solution for $\theta$ that minimizes the least square sum $\Vert \mathbf{A\theta}-\hat{\mathbf{b}} \Vert_2^2$ that can be obtained as:
 \begin{align}
 \label{eq:sls}
 \mathbf{A}^T\mathbf{A}\hat{\mathbf{\theta}}=
 \mathbf{A}^T\hat{\mathbf{b}}, \implies 
\hat{\mathbf{\theta}}=(\mathbf{A}^T\mathbf{A})^{-1}\mathbf{A}^T\hat{\mathbf{b}}.
\end{align}
The solution can be represented via Pseudo-Inverse as:
 \begin{align}
 \label{eq:re_sls}
\hat{\mathbf{\theta}}=\mathbf{A}^{\dagger}\hat{\mathbf{b}}.
\end{align}
The position estimate can be obtained from the first and second entries of $\hat{\mathbf{\theta}}$ as:
$\hat{\mathbf{p}}=\begin{bmatrix}[\hat{\theta}]_1 & [\hat{\theta}]_2\end{bmatrix}^T.$
where $[\hat{\theta}]_1=\hat{x}$, and $[\hat{\theta}]_2=\hat{y}$.
To determine the distribution of $\hat{\mathbf{p}}$, lets define $\hat{\mathbf{A}}^{\dagger}$ as ${\mathbf{A}}^{\dagger}$ with dimensions of $2*L$. Then the extracted estimated coordinates $\hat{\mathbf{p}}$ can be written based on eq. \ref{eq:re_sls} with an addition of the uncertainty as:
\begin{align}
\hat{\mathbf{p}}=\hat{{\mathbf{A}}}^{\dagger}\mathbf{b}+ \hat{{\mathbf{A}}}^{\dagger}\mathbf{n}
\end{align} 

\subsection{Hypothesis Testing}
At this stage, we need to classify the estimated position at the BS into a legitimate vehicle and a malicious vehicle.
Assuming $\mathbf{x}_A$ represents the vector of actual coordinates for a legitimate node and $\mathbf{x}_E$ represents the vector for a malicious node, we define $\mathcal{H}_0$ as the null hypothesis, indicating that the transmitter is the legitimate node, and $\mathcal{H}_1$ as the alternate hypothesis, suggesting that the transmitter is the malicious node. Then test statistics can be defined as:

 \begin{align}
    \text{TS}=\Vert \left( \hat{\mathbf{p}}-\hat{
 \mathbf{p}}_A \right)\Vert_2.
 \end{align}
 where $\hat{\mathbf{p}}_A $ is the ground truth provided by ML model.
The binary hypothesis test can be defined as
\begin{equation}
	\label{eq:H0H1}
	 \begin{cases} \mathcal{H}_0 (\text{no impersonation}): & \text{TS}=\Vert \left( \hat{\mathbf{p}}-\hat{
 \mathbf{p}}_A \right) \Vert_2 < \epsilon_{th} \\ 
                  \mathcal{H}_1 (\text{impersonation}): & \text{TS}=\Vert  \left( \hat{\mathbf{p}}-\hat{
 \mathbf{p}}_A \right) \Vert_2 > \epsilon_{th} \end{cases},
\end{equation}
where $\epsilon_{th}$ is a predetermined threshold. The binary hypothesis testing could also be defined as follows:
\begin{align} 
\label{eq:bht}
\text{TS} \gtrless_{\mathcal{H}_1}^{\mathcal{H}_0} {\epsilon_{th}}.
\end{align}

\subsection{ML Models for Mobility Tracking}
\begin{figure}[ht!]
    \centering
    \includegraphics[scale=0.3]{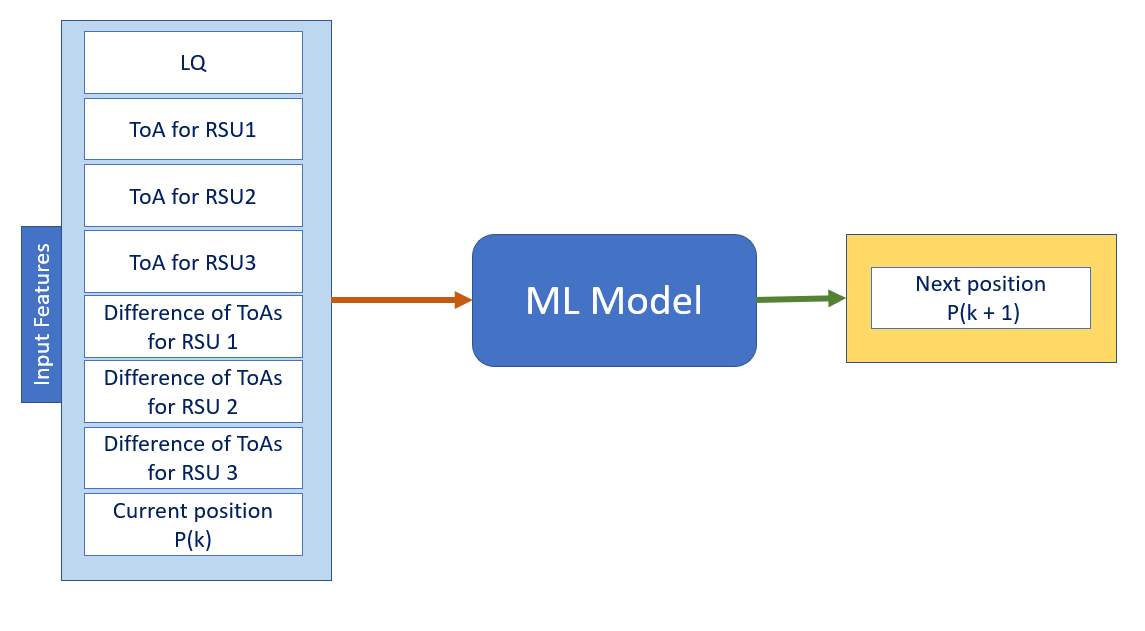}
    \caption{Model Structure}
    \label{fig:ML h}
\end{figure}

Mobility is a major challenge in VCNs. Typically, in PLA, one needs to get the ground truth information of the legitimate node in advance. In this work, to track the mobility pattern of the legitimate vehicle or get information about the ground truth of the legitimate vehicle, we employ the ML model(s). 
Generally, mobility models generate mobility traces by continuously predicting the next locations of the nodes. Such a location prediction process is basically a regression problem.  Hence, in this work, we use support vector regression (SVR) and decision tree (DT) as ML model(s). To train our models, we use eight input features: link quality (LQ), three TOAs and their three differences at the corresponding RSUs, and the current position of the vehicle to train our model. Our ML model(s) structure is shown in Fig. \ref{fig:ML h}. Note that we do not need extra efforts or estimation mechanisms to acquire these features as they are already available to BS.  
SVR is based on support vector machines (SVMs), which use a distinctive method for handling anticipated values that involves establishing a tolerance margin and predicts continuous output values differently than prior regression techniques. In this work, the trained SVR is used to forecast the output values for the x- and y-coordinates. 
Due to its decreased sensitivity to outliers and capacity to handle datasets with high-dimensional features, SVR is considered superior to other regression techniques.
On the other hand,  the decision tree (DT) regression approach divides the input data into smaller subgroups according to the values of the input features in order to forecast the outcome variable. DT can make precise, comprehensible forecasts.
\vspace{-0.2cm}
\section{Simulation}
\label{sec:simulation}
\subsection{Setup}
To evaluate the performance of our authentication scheme, we use MATLAB. Unless stated otherwise, the simulation parameters are listed in TABLE \ref{para}. We consider a long linear road of size $3000m \times 20m$. We deploy RSUs at both sides of the road at fixed positions, i.e., we separate any two RSUs at each roadside by $ 300$m. We assume RSUs are in LoS and with a distance from the transmitter vehicle of less than $400$m. Furthermore, we consider both vehicles moving at a certain speed. We assume that the malicious vehicle is smart enough and exactly following the legitimate node in speed and direction so that she can enhance the chances of missed detection. We also implement a scheme from the literature as a baseline scheme \cite{abdelaziz2018beyond} that uses the angle of arrival for authentication. It is implemented in MATLAB with the assumption that the actual ground truths of a moving legitimate vehicle are already available at BS. This assumption is taken because there is no explicit mechanism provided in \cite{abdelaziz2018beyond} to track or acquire the ground truths for a moving legitimate vehicle. 

\begin{table}[h!]
\centering
\begin{tabular}{| c | c |}
\hline
Parameter(s) & Value  \\
\hline
 Road Dimensions (Length and Width) & $3000$ m and $20$ m \\
Legitimate node position & [1,10]  \\
Malicious node position & [0,10]  \\
Speed of the vehicle & 1m/s  \\
Frequency & $18 \times 10^8 \mathrm{Hz}$  \\
path loss exponent  & 2 \\
Transmission power & $100 \mathrm{mW}$\\
Noise power & $\frac{P}{\mathrm{LQ}}$ \\
\hline
\end{tabular}
\caption{Monte Carlo Simulation Parameters}
\label{para}
\end{table}

\subsubsection {Dataset}
We generated a dataset for input features size of $9\times3.15*10^5$ and output labels of size $2\times3.15*10^5$. We randomly deploy 100 RSUs in a 2D region of size $5$km$\times$ $5$Km, define the starting position of the legitimate vehicle at a random position, and select the closest RSUs in LoS, which are under a predefined range, i.e., $400$m. Next, we vary the LQ in the unit step from 0dB to 20 dB, and for each LQ, We vary the speed of the legitimate vehicle from $0-33$mps (0-120kmph) randomly according to a uniform distribution. We then record the ToA for the selected three RSUs along with their differences (i.e., ToA in the previous slot subtracted from ToA in the current slot) and the extracted coordinate in $k$-th slot.  

\subsubsection{ML Models Configurations}
We use sequential minimal optimization (SMO) as a solver with loss function as MSE, given in Eq. \ref{eq:MSE} for SVR and least-square solver for DT. Note that these are default solvers used by MATLAB. We randomly divided the whole dataset into 0.7 and 0.3 segments for training and testing, respectively.

\subsection {Performance Evaluation Metrics}

\subsubsection{Analytical Model}
The performance metrics for the analytical model we adopt in this paper (i.e., hypothesis testing) are two error probabilities: the probability of false alarm $\text{P}_{\text{fa}}$ and the probability of missed detection $\text{P}_{\text{md}}$.  The  probability of false alarm is defined as the probability of incorrectly classifying a legitimate node as malicious during the binary hypothesis test, which can be expressed as: 
\begin{align}
    \text{P}_{\text{fa}}=\text{P}[\text{TS}\mid \mathcal{H}_0 \geq \epsilon_{th}]=\int_{\epsilon_{th}}^\infty f_{\text{TS}\mid \mathcal{H}_0}(\text{ts}\mid h_0)d_{\text{ts}\mid h_0}
\end{align}
 The probability of missed detection is the probability of incorrectly classifying a malicious node as legitimate during the binary hypothesis test, which can be expressed as:
\begin{align}
\label{eq:pmd}
    \text{P}_{\text{md}}=\text{P}[\text{TS}\mid \mathcal{H}_1 \leq \epsilon_{th}]=\int_{0}^{\epsilon_{th}} f_{\text{TS}\mid \mathcal{H}_1}(\text{ts}\mid h_1)d_{\text{ts}\mid h_1}
\end{align}
Note that the probability density functions ($f_{\text{TS}\mid \mathcal{H}_0}(\text{ts}\mid h_0)$ and $f_{\text{TS}\mid \mathcal{H}_1}(\text{ts}\mid h_1)$) are very challenging to find. We believe this requires dedicated long efforts to find out the nature of the conditional events ($\text{TS}\mid \mathcal{H}_0$ and $\text{TS}\mid \mathcal{H}_1$) and thus their density functions due to unknown uncertainty in the ground truths provided by the ML model and inherent uncertainty in the test statistics. Therefore, we compute these probabilities empirically in the simulations.

\subsubsection{ML Model}
Mean squared error (MSE), mean absolute error (MAE), and coefficient of determination $\text{R}^{2}$ are used as performance metrics to evaluate the performance of both SVR and DT models. 

MAE measures how far apart the expected and actual values are. The MAE can be expressed as:
\begin{equation}
    \text{MAE} = \frac{1}{2n}\sum_{i=1}^{n} \Vert \mathbf{p}_A^i - \hat{\mathbf{p}}_A^i \Vert_1,
\end{equation}

\noindent where $\mathbf{p}_A^i $ and $\hat{\mathbf{p}_A^i }$ stand for the $i-th$ observation's actual value and predicted value, respectively, while $n$ indicates the total number of observations and $\Vert.\Vert_1$ denotes $l_1$ norm operation.

MSE represents the difference between real and anticipated values, which can be mathematically defined as:
\begin{equation}\label{eq:MSE}
    \text{MSE} = \frac{1}{2n}\sum_{i=1}^{n}\Vert  \mathbf{p}_A^i - \hat{\mathbf{p}}_A^i \Vert_2^2,
\end{equation}
where $\Vert.\Vert_2$ denotes $l_2$ norm operation
\vspace{-0.1cm}
Finally, the coefficient of determination $\text{R}^2$ expresses how much of the variation in the dependent variable can be predicted from the independent variables. An $\text{R}^2$ value close to 1 indicates better performance, whereas an $\text{R}^2$ value significantly close to 0 indicates worse performance. The following defines the equation for $\text{R}^2$:
\begin{equation}
\text{R}^2 = 1 - \frac{\text{SS}_{\text{res}}}{\text{SS}_{\text{tot}}}
\end{equation}
where ${\text{SS}_\text{tot}}$ is the total sum of squares and $\text{SS}_\text{res}$ is the sum of squared residuals.

\subsection{Results}

\subsubsection{Error behavior against link quality (LQ)} 
\begin{figure}[ht!]
    \centering
    \includegraphics[scale=0.5]{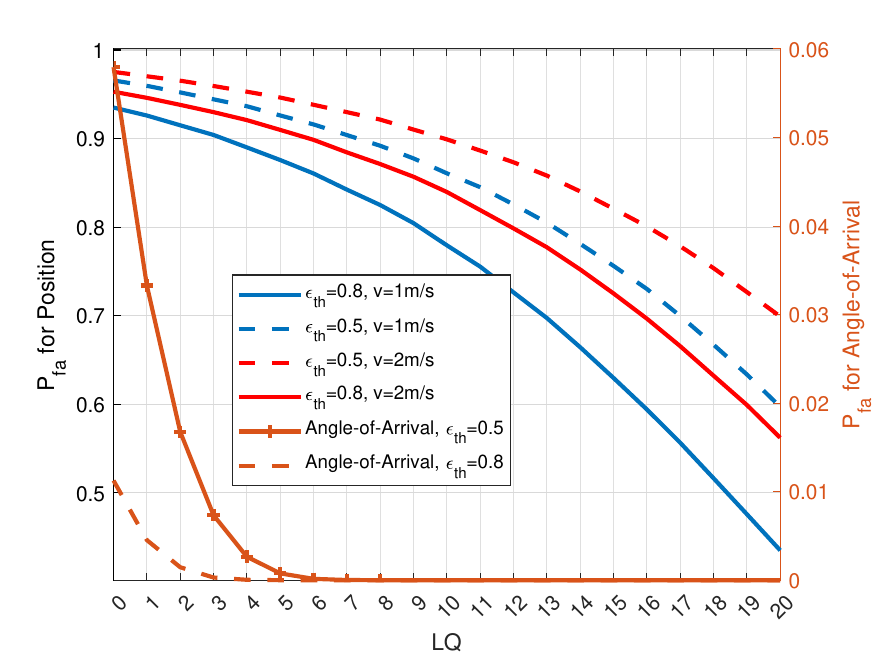}
    \caption{$\text{P}_{\text{fa}}$ vs LQ[dB]}
    \label{fig:pfaVSlqeps}
\end{figure}

\begin{figure}[ht!]
    \centering
    \includegraphics[scale=0.5]{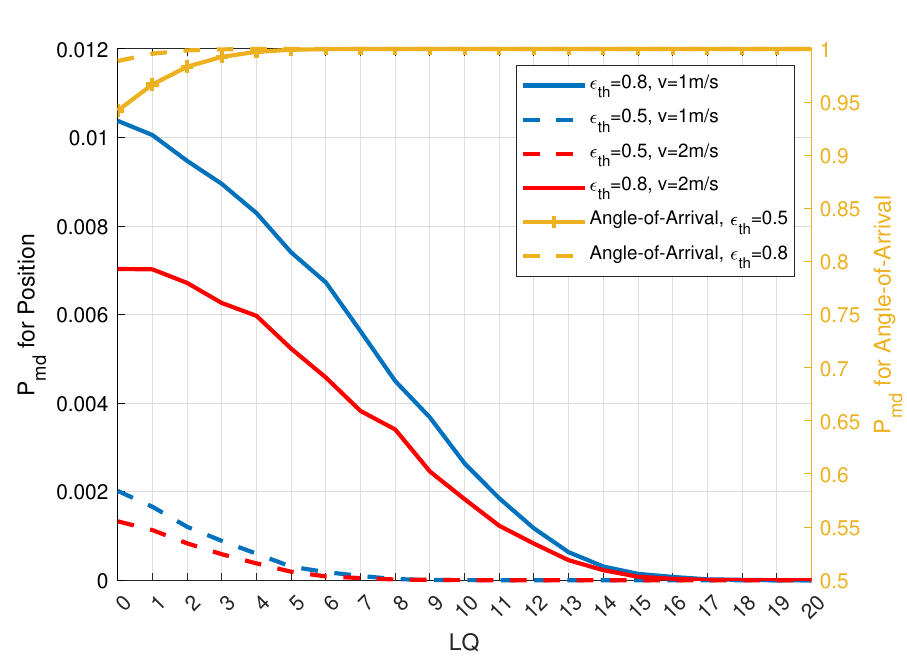}
    \caption{$\text{P}_{\text{md}}$ vs LQ[dB]}
    \label{fig:pmdVSlq}
\end{figure}

In this section we evaluate the performance of both error probabilities against LQ. We define LQ as the ratio of transmit power and noise power. Typically, to measure the LQ, a ratio of received power and noise power is taken but in our case due to multiple receivers (RSUs) a common variable is the ratio of transmit power and noise power among them, therefore, we redefine LQ as per our system model.  We sweep the LQ parameter from the 0 dB to 20 dB range in Figures \ref{fig:pfaVSlqeps}, \ref{fig:pmdVSlq}, and keep $1$m separation between legitimate and malicious vehicles. We observe as the LQ enhances both errors decrease for our proposed scheme.
On the other hand, if the design parameters of test statics, i.e., $\epsilon_{\text{th}}$  increases then a decrease in the probability of false alarm but an increase in the missed detection can be observed. We also investigate the impact of the velocity of the transmitter vehicles on the error probabilities. We notice that the increase in velocity has a positive impact (i.e., decreases in error) on the probability of missed detection  but a slight negative impact on the probability of false alarm. We also observe that the baseline scheme provides a very low false alarm for the same set of parameters but collapses on the probability of the missed detection which is an important or critical probability. The collapse (increase in error with the increase in LQ) of a fingerprint occurs when the fingerprints of both nodes are too close and the proposed scheme is unable to differentiate them.

\subsubsection{Receiver Operating Characteristic (ROC) curve}
\begin{figure}[ht!]
    \centering
    \includegraphics[scale=0.5]{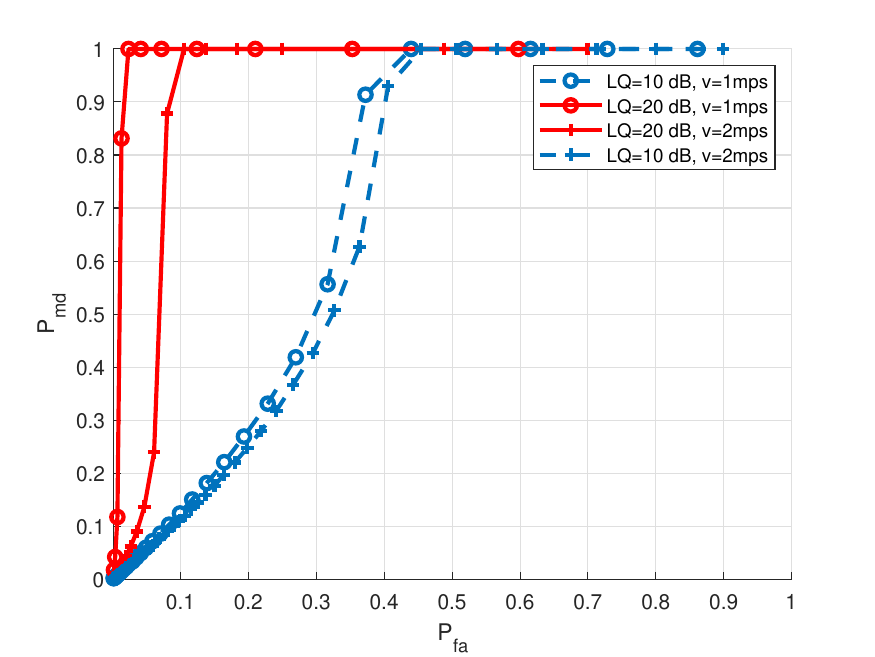}
    \caption{ROC: $\text{P}_\text{d}$ vs $\text{P}_{\text{fa}}$}
    \label{fig:ROC}
\end{figure}

ROC curves provide a comprehensive overview of our model, which allows us to evaluate the performance of our authentication scheme w.r.t. both errors. It shows the relationship between the detection rate (true positive) and the false alarm rate (false positive). ROC is generated by varying the threshold $\epsilon_{t h}$ over a long range and then for every single value of $\epsilon_{t h}$, both errors are recorded in arrays, and then plotted against each other. In Fig. \ref{fig:ROC}, we observe that the LQ has a positive impact on the detection rate ($P_d$ = $1-$ $P_{m d}$), the increase in LQ enhances the detection rate enhances. One can also see the impact of the speed of vehicles on the detection rate. Overall, speed has a negative impact on the performance of the proposed scheme. 

\subsubsection {ML Models Performance Results}

We present the performance of the above-mentioned two ML models in TALBE \ref{perf} based on the test dataset of size $9 \times 10^5$ generated as per the considered mobility pattern. 
\begin{table}[h]
\centering
\begin{tabular}{|c|c|c|c|c|}
\hline
\textbf{Model} & \textbf{RMSE} & \textbf{MSE} & \textbf{MAE} & \textbf{R$^2$} \\ \hline
DT      & 0.40837      & 0.166765         & 0.227203         & 0.498604           \\ \hline
SVR      & 0.27982     & 0.078298        & 0.169555        & 0.764588            \\ \hline
\end{tabular}
\caption{Comparison between DT and SVR 
based on RMSE MSE, MAE, and $R^2$}
\label{perf}
\end{table}
 Note that root MSE (RMSE) is the square root of MSE. Overall, SVR model's predictions appear to have better performance than DT model. As a result, it is more suitable for this study based on RMSE, MSE, MAE, and $R^2$ measurements.



\section{Conclusion}
\label{sec:conclusion}
In this paper, we developed a novel hybrid physical layer authentication scheme with ML for V2I communication that uses the location of the transmitter as a fingerprint. We consider a dynamic environment for transmitter vehicles where nodes are mobile. The proposed authentication scheme is tested against various parameters of the system, i.e. speed of the vehicles, link quality, and controlled parameter threshold. The performance is also compared with a baseline scheme that exploits the angle of arrival as a device fingerprint. We showed that position is a strong candidate feature for PLA, where one can achieve high detection rates even at with low link quality. 

This work can be extended by studying our proposed scheme with more realistic and non-linear mobility models, and trying more ML models for better accuracy. Similarly, this work can be extended by incorporating multiple legitimate and malicious vehicles, which is novel in the context of PLA in VCNs.   

\bibliographystyle{IEEEtran}
\bibliography{references}

\vfill\break

\end{document}